\documentclass[reprint, aps, prab, amsmath, amssymb, unsortedaddress]{revtex4-2} 
\usepackage{graphicx} 
\usepackage[caption=false]{subfig} 
\usepackage[hidelinks]{hyperref}
\usepackage{xcolor}
\usepackage{lineno}
\usepackage{siunitx}

\newcommand{\eq}[1]{(\ref{#1})}

\newcommand{\E}{\ensuremath{\mathbf{E}}}

\newcommand{\permit}{\varepsilon_0}

\begin{document}

\title{Tolerances to driver-witness misalignment in a quasilinear plasma wakefield accelerator}
%\author{T. C. Wilson$^{1,2}$, J. Farmer$^3$, A. Pukhov$^{4}$}

\author{T. C. Wilson}
\affiliation{Institut f\"{u}r Theoretische Physik I, Heinrich Heine Universit\"{a}t D\"{u}sseldorf, 40225 Germany}

\author{J. Farmer}
\affiliation{Max-Planck-Institut f\"{u}r Physik, 85748 Germany}

\author{K. Lotov}
\affiliation{Budker Institute of Nuclear Physics, 630090 Novosibirsk, Russia}
\affiliation{Novosibirsk State University, 630090 Novosibirsk, Russia}

\author{A. Pukhov}
\affiliation{Institut f\"{u}r Theoretische Physik I, Heinrich Heine Universit\"{a}t D\"{u}sseldorf, 40225 Germany}

\begin{abstract}
Plasma-based accelerators offer high accelerating gradients and scalability through staging or long plasma sources, which makes them good candidates for future accelerator and collider concepts. Proton-driven accelerators in particular have the potential to bring particles to high energy in a single stage. 
In the quasilinear regime - where the plasma wake is only partially evacuated - a witness bunch of electrons drives a cavitated wake, which acts to preserve the emittance of the portion of the witness inside this self-blowout. In the case of a misalignment between the driver and witness, this behaviour can persist, but its effectiveness is reduced. 
In this paper, we study transverse witness dynamics in this regime, and develop analytical models to describe the witness motion, and develop a metric to estimate emittance preservation based on a single parameter which estimates the density of the witness after phase mixing.
Particle in cell simulations using the AWAKE Run 2c baseline parameters show excellent agreement with the predictive models developed. This work allows alignment constraints to be set both for the AWAKE experiment and other wakefield acceleration schemes operating in the quasilinear regime.
\end{abstract}

\maketitle
\section{Introduction}
Prospective designs of future particle accelerators look to push the energy frontier forward. In order to do this, the high acceleration gradients of plasma-based acceleraton are attractive \cite{Litos16, Zhang25}. While the accelerating gradient of conventional radiofrequency (RF) cavity machines are limited to the 10s of \SI{}{\mega\electronvolt\per\metre} due to RF breakdown, plasmas can support high accelerating gradients at the \SI{}{\giga\electronvolt\per\metre} level, and hence have the potential to reduce the real-estate footprint of future accelerator facilities \cite{Litos16, Foster23, Farmer24, ALiVE25}.

In order to use plasma as an accelerator, high electric fields must be set up by separating the electrons and ions. This requires a suitable driver, either an intense short-pulse laser \cite{Tajima79} or an energetic particle beam \cite{Chen85}. 
This driver displaces plasma electrons as it propagates, while the much heavier ions move comparatively much less. This separation of charges induces electric fields which cause the plasma electrons to oscillate, and a travelling electron density wave with a phase velocity equal to the driver velocity follows the driver. This is a plasma wake, into which a `witness' bunch of electrons may be injected and accelerated to high energy. 

As the driver propagates through the plasma, energy is gradually transferred to the wake, and ultimately to the witness. If the driver propagates sufficiently far, it will lose a significant fraction of its energy and the wake will collapse, making acceleration no longer possible. Collider-relevant plasma accelerators will need to be hundreds to thousands of metres in length, which exposes them to the possibility of driver depletion.

Two main schemes address this problem. 
Either, periodically replacing the spent driver in order to continue the acceleration of the witness, in a process called `staging'. 
Staging requires both the extraction of the spent driver, and injection of a new one, whilst simultaneously preserving the witness \cite{Lindstrom21}.

Alternatively, by choosing a driver with a high enough energy, the depletion length is guaranteed to be longer than the accelerator. Proton beams are the only practical choice for such a scheme \cite{Caldwell09, Gschwendtner16}, with their extremely high mass allowing them to be accelerated with low synchrotron losses in existing circular collider infrastructure.

We consider here the latter, a proton-driven plasma wakefield accelerator. 
Proton beams are the only currently-available driver which can reach the energy frontier in a single stage.

The AWAKE experiment \cite{Gschwendtner16, Gschwendtner22} is a proof-of-principle accelerator using the \SI{400}{\giga\electronvolt} proton beam from the CERN Super Proton Synchrotron (SPS) to accelerate electrons in plasma. The SPS beam is around \SI{5}{\centi\metre} in length, and the plasma is quite sparse, having a plasma wavelength of $\sim\SI{1}{\milli\metre}$, thus the beam spans many plasma periods. The beam is initially too long to effectively drive a strong wake for acceleration, but nontheless the periodic focussing and defocussing action causes it to undergo self-modulation. The beam modulates over several metres into a train of microbunches \cite{Kumar10, AWAKE18} with periodicity on the scale of the plasma wavelength. The modulated beam is then suitable for resonant wakefield excitation.

As the proton beam modulates, the phase of the wake shifts. It is therefore undesireable to attempt acceleration until the wake has stabilised \cite{Lotov14}. The AWAKE Run 2c experiment will separate the modulation stage from the acceleration stage physically, in two sequential plasma sources. This allows the proton beam to completely self-modulate in the first, and then electrons to be injected into a stable wake in the second. Each stage will be \SI{10}{\metre} long, with a gap of \SI{1}{\metre} to allow for injection \cite{Gschwendtner22}.

Building upon the proton driven accelerator concept, the ALiVE scheme \cite{ALiVE25} proposes to use a single short proton bunch \cite{Willeke24, flowerdew:ipac2025-weps028} to drive the wakefield, thus avoiding the need for modulation entirely.

\begin{figure}
\centering
\includegraphics[width=\linewidth]{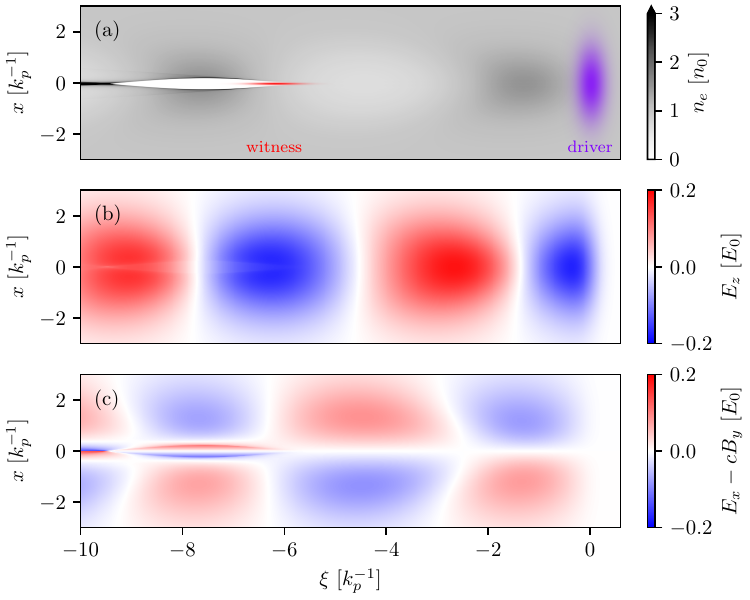}
\caption{Schematic view of the simulation setup. For an aligned driver-witness setup we have (a) the plasma (grey), with driver (purple) and witness (red) moving left-to-right in series. (b) The longitudinal fields, showing the beamloading due to the witness, and (c) the transverse force seen by a relativistic particle. We see that the witness blowout creates a region of focusing fields within the larger wake. Fields are normalised to the wavebreaking field $E_0 = m_ee\omega_p/e$. Here $\xi$ is the longitudinal coordinate relative to the driver centroid.}
\label{toymodel}
\end{figure}

When accelerating electrons, it is typical to operate in the blowout regime. This is where the driver expels all plasma electrons behind it, leaving only the background ions. This is preferable not only because it produces the strongest possible accelerating fields for a given plasma density, but also because the focussing fields on the bunch are linear. By matching the beta function of the witness to the focussing field of the plasma, the witness emittance can be preserved \cite{Rosenzweig91}. 
However, proton-driven wakefields tend to operate in the quasilinear regime, where the wake is not fully evacuated, and so we cannot expect to preserve the witness emittance completely. This being said, the witness itself also drives wakefields and, when its charge density is sufficiently high, can drive its own blowout \cite{Olsen18}. This affords some degree of emittance control. A schematic view of this is shown in Figure \ref{toymodel},  where we depict a short driver, rather than a train of microbunches. 

For an externally injected witness, we consider the effect of imperfect transverse alignment between the driver and witness at injection. This will be present to some degree in any real experiment, and has been studied for the blowout case \cite{Adli25}, where tight tolerances were found. 
In the context of a quasilinear wake, we look at the effect a transverse misalignment has on the final beam properties, and what experimental tolerances might be extracted from this.

\section{Simulation setup}
We performed simulations using the 3D quasistatic PIC codes QV3D \cite{Pukhov16} and HiPACE++ \cite{Diederichs22}, and found excellent agreement between the two. The following parameters follow those of \cite{Olsen18}, where a similar study was conducted. We here present a more detailed study of emittance preservation in the presence of an initial misalignment.
We simulate a plasma of length \SI{10}{\metre} and radius \SI{1}{\milli\metre}, with a constant plasma density of $n_0=\SI{7e14}{\per\cubic\centi\metre}$, corresponding to a skin-depth of $1/k_p = \SI{200}{\micro\metre}$. The \SI{400}{\giga\eV} driver is modelled as a positively-charged, non-evolving Gaussian ellipsoid, with a radius of $\sigma_{r,p^+} = \SI{200}{\micro\metre}$, length $\sigma_{z,p^+}=\SI{40}{\micro\metre}$, and total charge $Q =\SI{2.34}{\nano\coulomb}$ ($n_b = 0.83n_0$), resulting in a peak accelerating gradient of $\sim\SI{500}{\mega\electronvolt\per\metre}$. These parameters were chosen to emulate the wakefields driven by the full microbunch train in AWAKE Run 2c \cite{Gschwendtner22}.

The electron witness bunch is also a Gaussian ellipsoid, with an injection energy of $\SI{150}{\mega\eV}$ ($\gamma_0\approx 294$) with $\SI{0.1}{\percent}$ energy spread, a longitudinal length of $\sigma_z =\SI{60}{\micro\metre}$, and a radial size chosen to satisfy to the blowout matching condition
\begin{equation}
\sigma_{\mathrm{matched}} = \left(\frac{2\epsilon_n^2}{k_p^2\gamma}\right)^{1/4},
\label{matched_radius}
\end{equation}
where $\epsilon_n$ is the normalised emittance, $\gamma$ is the Lorentz factor of the electron witness, and $c$ is the speed of light in vacuum. 
We choose to use the blowout matching condition despite the proton-driven wake being quasilinear, as we expect most of the bunch charge to sit in a self-generated blowout.

As the plasma streams past the witness, the blowout takes some finite time to form, and this time depends on the charge density \cite{Olsen18, Farmer21}.
We vary the witness emittance from $\epsilon_n = 2$ -- \SI{16}{\micro\metre}, corresponding to a range of radial sizes $\sigma_r = 5.76$ -- \SI{16.3}{\micro\metre}, and hence a range of peak charge densities. Beam charge is also varied from 100 -- \SI{400}{\pico\coulomb}, and the transverse offset between the driver and the witness is varied from 0--\SI{20}{\micro\metre}. 
The longitudinal position of the witness in the wake is chosen so as to minimise energy spread, and varies linearly between 4.1 -- \SI{3.8}{\pico\second} ($6.2$ -- $5.7/k_p$) depending on the witness charge (high -- low). 

\section{Results and discussion}
A schematic view of the simulation setup is shown in Fig. \ref{toymodel}, where we illustrate the initial configuration for the perfectly aligned witness case. Fig. \ref{toymodel}(a) shows the relative densities of the plasma, driver, and witness, where we see immediately that the wake is not fully evacuated by the low-density driver, allowing the higher-density witness to drive its own blowout within the larger wake. We also see the effect of beamloading in Fig. \ref{toymodel}(b), situated over the witness by the fact the accelerating fields are reduced (paler in colour). Finally the transverse fields are shown in Fig. \ref{toymodel}(c), where the quasilinear wake fields are disturbed by the small blowout driven by the witness. We notice that the witness-blowout projects a focussing region even when located in the defocussing phase of the encompassing quasilinear wake.

\subsection{Key results}
\begin{figure}
\centering
\includegraphics[width=\linewidth]{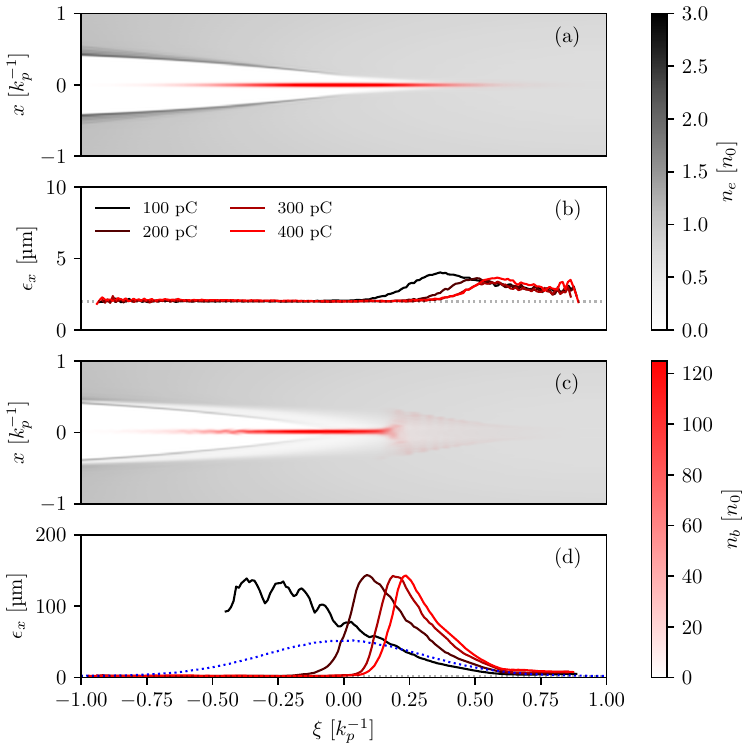}
\caption{Snapshots of the \SI{400}{\pico\coulomb}, \SI{2}{\micro\metre} emittance bunch for (a) no offset and (c) \SI{20}{\micro\metre} offset after \SI{10}{\metre}, and comparative slice emittances for all charge cases 100 -- \SI{400}{\pico\coulomb}, again for (b) no offset and (d) a \SI{20}{\micro\metre} offset. The initial charge density profile of all beams is illustrated by the blue dotted line in (d). Here $\xi$ is the longitudinal coordinate relative to the witness centroid.} 
\label{emittance}
\end{figure}
The simulations were allowed to evolve over \SI{10}{\metre}, and in this time, projected energy spread increases to 5 -- 10\% depending on charge (low -- high). Energy gain varies inversely with bunch charge, falling in the range 2 -- \SI{4}{\giga\electronvolt} (high -- low). Both final energy and projected energy spread are mostly independent of the initial offset.

Figure \ref{emittance} shows final density maps and a summary of final slice emittances for different configurations. 
Figure \ref{emittance}(a) shows a snapshot after 10m of acceleration for the case of no initial offset. The \SI{400}{\pico\coulomb}, \SI{2}{\micro\metre} case is shown, but this picture is typical; the head of the witness sits in the quasilinear wake generated by the driver, while the tail sits in the self-blowout.

Figure \ref{emittance}(b) shows the slice emittance for this case (red line), and all other witness charges with no initial offset and \SI{2}{\micro\metre} initial emittance. 
Using the matched radius ensures that for the portion of the witness which sits in a blowout, there is little change in emittance, and this is reflected in the slice emittances. We see that despite no visible degradation of the beam head in Fig. \ref{emittance}(a), there is still some emittance growth, which drops off as the local plasma electron density approaches zero. The higher the charge of the bunch, the earlier a blowout forms, and emittance preservation occurs for a larger fraction of the beam. 
Overall, the projected emittance growth varies from $\epsilon_x= \SI{2}{\micro\metre}$ to 2.1 -- \SI{2.5}{\micro\metre} depending on charge, with higher charge seeing less growth.

Figure \ref{emittance}(c) shows the equivalent case to Fig. \ref{emittance}(a), but now with a $0.1k_p^{-1}$ (\SI{20}{\micro\metre}) initial transverse offset. We see that after \SI{10}{\metre} the witness has taken on a distinctive profile, where the head of the bunch has become smeared out beyond the range of the initial offset. This occurs as the low-density head is not matched to the focussing fields in the quasilinear wake, and so the beam slices oscillate and undergo phase-mixing, resulting in an increased radius and emittance. Some length back from the head of the beam, at around $\xi=0.2k_p^{-1}$, a blowout forms and the tail of the bunch remains tightly focussed, oscillating coherently with emittance preservation. The blowout forms further back along the bunch compared to the zero-offset case, and this distance increases with increasing offset.

Figure \ref{emittance}(d) shows the final slice emittances for bunches with a \SI{20}{\micro\metre} offset, and \SI{2}{\micro\metre} initial emittance. In the lowest-charge case (black line) the dilution of the bunch head means the transverse wake is no longer strong enough to keep the witness focussed, and the entire length of the bunch experiences massive emittance growth, with a final projected emittance of $\epsilon_x = \SI{80}{\micro\metre}$.
However, at higher witness charge, emittance can still be partially preserved as the increasing density of the witness still allows a blowout to form, which protects progressively more of the bunch, with the \SI{400}{\pico\coulomb} case having a final projected emittance of $\epsilon_x = \SI{25}{\micro\metre}$.

\subsection{Witness betatron motion}
\begin{figure}
\centering
\includegraphics[width=\linewidth]{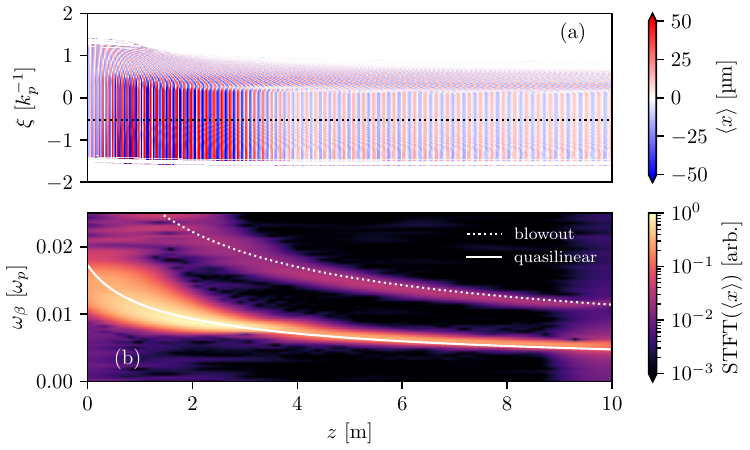}
\caption{(a) Waterfall plot of the bunch transverse centroid position over its length $\xi$ and the propagation coordinate $z$ and, (b) short-time Fourier transform, showing the propagation distance-resolved oscillation frequency at the point $\xi = -0.5k_p$ in the witness, indicated by the dotted line in (a). The dotted white line shows the blowout betatron frequency $\omega_p/\sqrt{2\gamma}$, and the solid white line shows the numerically calculated quasilinear betatron frequency.}
\label{STFT}
\end{figure} 
In Fig. \ref{STFT}(a) we plot the transverse centroid position of the \SI{400}{\pico\coulomb} bunch with a \SI{20}{\micro\metre} initial offset, as shown in Fig. \ref{emittance}(c), bunch along both the length of the bunch itself ($\xi$) and along the propagation direction ($z$). We see there is a visible indicator of the regime change at around $\xi\approx0.5/k_p$ between the head of the bunch, undergoing phase-mixing, and the rear of the bunch, undergoing coherent betatron oscillations. 

Figure \ref{STFT}(b) shows the betatron frequency of the witness as a function of propagation distance. This is obtained by choosing a fixed $\xi$, and taking the witness transverse centroid position along $z$ at this point, indicated by the dotted line in Fig. \ref{STFT}(a). Applying a short-time Fourier transform (STFT) on this signal yields the betatron wavenumber $k_\beta$ as a function of propagation distance $z$.
We can then approximate the temporal frequency as $\omega_\beta = ck_\beta$, representing the effective betatron frequency at the chosen $\xi$. We see that there are two frequency components to the signal, with the low-frequency component dominant.

This may be explained by analogy. Consider a ball (the witness) placed in a basket (its self-blowout), and the basket is moved back and forth at low frequency (the oscillations of the driver head in the quasilinear wake). However, the ball's position within the basket is not fixed, its own intertia will cause it to undergo high frequency oscillations depending on the shape of the basket (the the transverse potential of the self-blowout). 
When the frequency of the ball's motion is measured, there will be two components; the low-frequency motion of the basket, and the high-frequency motion of the ball within the basket. As the amplitude of the slow motion is much greater than that of the fast motion, the low-frequency component will have the stronger signal. 

We may retrieve an estimate the low-frequency component by modelling the quasilinear wakefield potential via Poisson's equation 
\begin{equation}
\nabla^2\varphi = -\frac{\rho}{\permit},
\end{equation}
where $\varphi$ is the wakefield potential, $\rho$ is the plasma charge density, and $\permit$ is the permitivity of free space.
By assuming separable solutions $\varphi = \varphi_z \varphi_\perp$, $\rho =\rho_z\rho_\perp$, the longitudinal components may be eliminated by assuming a sinusoidal density modulation $\rho = \rho_\perp \cos(k_pz)$, in line with the quasilinear approximation. The corresponding potential is therefore $\varphi = -\varphi_\perp \cos(k_pz)$. We hence find that the transverse wake potential takes the form of a screened Poisson equation
\begin{equation}
\label{screenedpoisson}
\left(\nabla_\perp^2 - k_p^2\right)\varphi_\perp = \frac{\rho_\perp}{\permit},
\end{equation}
where $\nabla_\perp^2$ is the transverse part of the Laplacian. This is an intuitive result, as any space charge fields in plasma are naturally screened by the background plasma electrons.

For our setup, we find the unloaded wake density falls to $n_e \approx 0.67 n_0$. We may solve \eq{screenedpoisson} numerically and then retrieve the fields from $\E=-\nabla\varphi$. We make the small-displacement approximation and take $E_\perp$ only to lowest order, so as to model the motion as a simple harmonic oscillator. 
We find that the the effective focussing field is reduced from \cite{Lotov04} \begin{equation}
E_{\perp, \mathrm{blowout}} \approx  \frac{n_0e}{2\epsilon_0},  
\label{wakefields}
\end{equation}
(where we take $B_\theta$ to be small) to $E_{\perp, \mathrm{quasilinear}} \approx 0.175 E_{\perp, \mathrm{blowout}}$. From this, we can approximate a reduced betatron frequency as a function of $z$ via the relation 
\begin{equation}
\omega_\beta = \sqrt{\frac{eE_\perp}{\gamma m_e}},
\label{betatron}
\end{equation}
where $\gamma$ is varying with $z$.
Comparing this to simulation, in Fig \ref{STFT}(b) we overlay our numerically-calculated low-frequency component (solid line), and the high-frequency component (dotted line) is the unscreened blowout betatron frequency (\eq{wakefields} in \eq{betatron}), and find them both to be in good agreement with the results.

\subsection{Parameter scan}
\begin{figure}
\centering
\includegraphics[width=\linewidth]{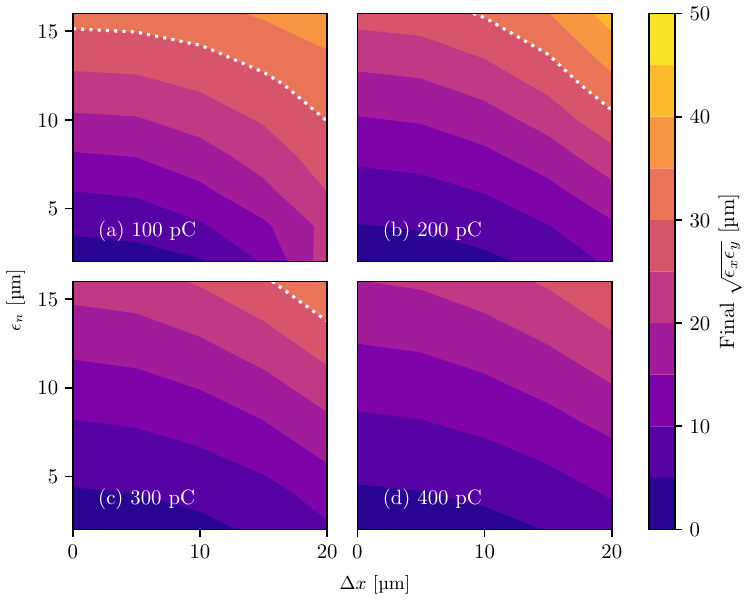}
\caption{Projected emittance after 10 metres. The geometric average of the $x$ and $y$-plane emittance is taken as the representative quantity while varying the initial emittance $\varepsilon_n$ and the offset $\Delta x$. Subplots (a) - (d) show increasing charge from 100  - 400 pC. The isobars for a \SI{30}{\micro\metre} projected emittance are shown as white dotted lines.}
\label{projected_emittance}
\end{figure}
Expanding the scope of our simulations, we can scan several parameters and look for patterns. Taking initial emittance in the range 2 -- \SI{16}{\micro\metre}, charge in 100 -- \SI{400}{\pico\coulomb} and offsets in 0 -- \SI{20}{\micro\metre}, we plot the resulting emittance as a function of these three variables in Fig. \ref{projected_emittance}.
We have the intuitive result that in general, higher charges results in less emittance growth across the board. This can be seen by observation, and for a visual aid we mark the isobar for a \SI{30}{\micro\metre} projected emittance on all figures. One can see that for increasing charge, this isobar moves further up, indicating emittance preservation for a wider range of parameters.

With an offset, the emittance in the plane of initial offset will be larger than that of the non-offset plane. While the emittance in the plane of the offset  (here $x$) grows considerably, as seen in Fig. \ref{emittance}(d), the emittance in the orthogonal plane (here $y$) remains on the order of a non-offset bunch e.g. Fig. \ref{emittance}(b). We use the geometric mean of the $x$ and $y$-emittances to get a representative quantity for the overall bunch quality, resulting in the more modest increases to the projected emittance recorded in Fig. \ref{projected_emittance}.

In the presence of an initial offset, we expect the head of the bunch to undergo rapid phase-mixing. This will cause the radius in the plane of the offset to increase proportional to the size of the offset. In order to quantify the results we propose to define an effective density of a phase-mixed witness
\begin{equation}
n_{b,\mathrm{eff.}} = \frac{Q}{e(2\pi)^{3/2}\sigma_z\sigma_y\sqrt{\sigma_x^2+\Delta x^2}}.
\end{equation}
This quantity encodes the charge, offset, and emittance (through the matched $\sigma_{(x,y)}$ \eq{matched_radius}), and is analagous to the peak density of a Gaussian bunch that has been spread out in one direction by a factor $\sqrt{1 + (\Delta x/\sigma_x)^2}$, giving a represenation of the power of a bunch to drive its own blowout. 

\begin{figure}
\centering
\includegraphics[width=\linewidth]{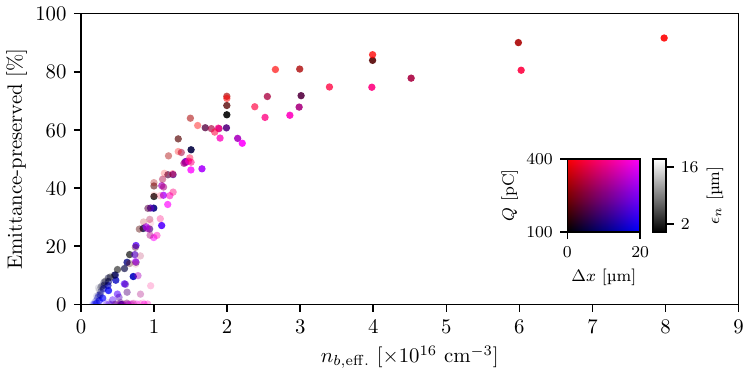}
\caption{Fraction of the bunch charge for which emittance is preserved (final emittance $<1.1$ times initial emittance). The effective peak density of the bunch is taken as $n_{b,\mathrm{eff.}} = Q/[e(2\pi)^{3/2}\sigma_z\sigma_y\sqrt{\sigma_x^2+\Delta x^2}]$. The point colour reflects the bunch parameters; charge (redness), offset (blueness) and initial emittance (alpha).}
\label{preserved_fraction}
\end{figure}
We now compute the preserved emittance fraction of each case by examining the slice statistics. We look at the slice emittance along the bunch, and choose some threshold emittance, under which we consider that slice to be `preserved'. Setting this threshold is arbitrary, we here choose a threshold of 10\% over the initial emittance. We then add up the charge of each slice under the threshold, and then compare this to the charge of the whole bunch. This gives us a `preserved fraction' for each case. 

The result of this procedure is plotted in Fig. \ref{preserved_fraction}, where we see a strong correlation between the preserved fraction and our effective density. 
Choosing a different threshold value changes the shape of the curve, lower values flatten it, and higher values steepen it, but the general trend is maintained.
There are a collection of points with low effective density that have little-to-no emittance preservation at all. This is not unexpected \cite{pwfa-farmer-selfmatching, pwfa-romeo-assistedmatching}, as it stands to reason that there will be some critical threshold below which no blowout can be driven at any offset and thus no emittance preservation.

We may use this information to determine tolerances. By nominating a desired final emittance, one can find the maximum offset for which the required performanc is achieved, give the initial bunch charge and emittance.
Additionally, we can extend this to estimate a pointing error. Assuming the bunch motion is sinusoidal, a transverse offset $\Delta x$ can be translated into an angle $\Delta\theta$ by considering the trajectory of a test particle as it crosses the axis,
\begin{equation}
\Delta\theta = \arctan(k_\beta \Delta x) \approx k_\beta \Delta x,
\end{equation}
where $k_\beta = \omega_\beta/c$, and we assume $\Delta\theta$ remains small. These results are encouraging if translated to angular misalignment, the largest offset considered here corresponds to an angular offset on the order of \SI{2}{\milli\radian}, which is large compared to currently-achievable performance. This arises due to the large focussing fields in plasma, such a small angle offset does not lead to large oscillations.

\section{Conclusion}
Transverse or angular misalignments when staging or injecting act to degrade beam quality, and so effective emittance control is a challenge that must be addressed for future plasma-based accelerators and colliders.

We find that plasma acceleration in the quasilinear regime exhibits unique transverse dynamics, which can be split into two modes; the slow betatron motion of particles in the quasilinear wake, and the fast betatron motion of those within the self-bubble. The fast motion is described well by existing theory, and we have developed an analytical model for the slow betatron motion, describing witness self-blowout motion in the case of an offset. The witness head oscillates, which leads to oscillation of the self-blowout, within which the witness oscillates at the textbook betatron frequency.

The time for the witness self-blowout to form depends on the charge density of the beam head, and greatly influences the degree of emittance preservation. We show that the final emittance depends on a single metric derived from the beam parameters and offset, which estimates the bunch charge denstiy after phase mixing. The predictive power of this model is tested by simulation and shows strong correlation.

Using the model, tolerances can easily be determined for a desired level of emittance control, and positional offsets can be translated into equivalent angular offsets. 
A beneficial property of plasma is that typical angular tolerances are relaxed compared to transverse misalignment tolerances, due to the strong focussing action of the plasma.

This work will be useful for the ongoing design of AWAKE Run 2c and beyond, and for any other acceleration scheme which operates in the quasilinear regime.

\section{Acknowledgments}
This work used computational resources of the National Energy Research Scientific Computing Center, which is supported by DOE DE-AC02-05CH11231.

The authors gratefully acknowledge the Gauss Centre for Supercomputing e.V. \cite{GCS} for funding this project (lpqed) by providing computing time through the John von Neumann Institute for Computing (NIC) on the GCS Supercomputer JUWELS-Booster at J\"{u}lich Supercomputing Centre (JSC).

% to generate from .bib
\bibliographystyle{unsrt}
\bibliography{bibliography}

% .bbl file contents begins here
%\begin{thebibliography}{10}

%\end{thebibliography}
% .bbl file contents ends here

\end{document}